\documentclass[12pt, letterpaper]{article}

\usepackage{amsmath,amssymb}
\usepackage{cite}
\usepackage{fancyhdr}
\usepackage[top=1in, bottom=1in, left=1in, right=1in]{geometry}
\usepackage[dvipdfm, dvips]{graphicx}
\usepackage{hyperref}

\numberwithin{equation}{section}

\begin{document}


\setcounter{page}{0}
\date{}

\lhead{}\chead{}\rhead{\footnotesize{RUNHETC-2011-12\\SCIPP-11-13}}\lfoot{}\cfoot{}\rfoot{}

\title{\textbf{Holographic Space-Time: The Takeaway \vspace{0.4cm}}}

\author{Tom Banks$^{1,2}$\vspace{0.7cm}\\
{\normalsize{$^1$NHETC and Department of Physics and Astronomy, Rutgers University,}}\\
{\normalsize{Piscataway, NJ 08854-8019, USA}}\vspace{0.2cm}\\
{\normalsize{$^2$SCIPP and Department of Physics, University of California,}}\\
{\normalsize{Santa Cruz, CA 95064-1077, USA}}}

\maketitle
\thispagestyle{fancy}

\begin{abstract}
\normalsize \noindent The theory of holographic space-time (HST)
generalizes both string theory and quantum field theory. It provides
a geometric rationale for supersymmetry (SUSY) and a formalism in
which super-Poincare invariance follows from Poincare invariance.
HST unifies particles and black holes, realizing both as excitations
of non-commutative geometrical variables on a holographic screen.
Compact extra dimensions are interpreted as finite dimensional
unitary representations of super-algebras, and have no moduli. Full
field theoretic Fock spaces, and continuous moduli are both emergent
phenomena of super-Poincare invariant limits in which the number of
holographic degrees of freedom goes to infinity. Finite radius de
Sitter (dS) spaces have no moduli, and break SUSY with a gravitino
mass scaling like $\Lambda^{1/4}$. We present a holographic theory of inflation
and fluctuations.  The inflaton field is an emergent concept, describing the geometry
of an underlying HST model, rather than ``a field associated with a microscopic string theory".
We argue that the phrase in quotes is meaningless in the HST formalism.
\end{abstract}


\newpage
\tableofcontents
\vspace{1cm}


\section{Introduction}

This note was written as a response to various people who asked me for a short
introduction to the principles of Holographic Space-Time (HST).  It includes descriptions of ideas whose details
have not yet appeared in print, as well as older work, which has been largely ignored.  
The note consists of a large number of short sections, each devoted to a key principle of the theory.

\section{Causal Diamonds and Observables}

In classical Lorentzian geometry, a {\it causal diamond} is a subset of d dimensional space-time on which an observer, traveling from a point P, to a point Q in the future of P, can do experiments.  It consists of the intersection between the interior of the backward light-cone from Q and the forward light cone from P.  The boundary of the diamond is a null surface. If we consider any foliation of space-time by space-like surfaces, each surface cuts the boundary in a $d-2$ dimensional space-like surface. The maximum, within a foliation and among all foliations, area $d-2$ surface is called the {\it holographic screen} of the diamond.

In quantum field theory (QFT), if one smears the fields with test functions of compact support in the diamond, one obtains a tensor factor $A(D)$ of the full operator algebra.  Its commutant, the set of operators commuting with it, consists of fields smeared with test functions whose support is space-like separated from the diamond.  If two diamonds intersect, their intersection contains a causal diamond of maximal area.  Fields smeared with functions of support in that maximal intersecting diamond, form a tensor factor $O(D_1 , D_2)$ in each of the individual diamond algebras $A(D_i) $.   The causal structure of space-time is thus embedded implicitly in the quantum operator algebra.  HST retains this structure.

In QFT, the operator algebras $A(D)$ are all infinite dimensional and depend both on the conformal field theory (CFT) and relevant perturbation of it, which defines the particular theory.  The Holographic Principle tells us that Quantum Gravity is dramatically simpler: {\it  the diamond operator algebras are all finite $N\times N$ matrix algebras, operating in a Hilbert space of dimension $e^{A/4}$, where $A$ is the area in Planck units, of the holographic screen (henceforth holoscreen)}\cite{rbwftb}.  This principle has more to recommend it than mathematical simplicity.  The Holographic Principle tells us that both the causal structure and conformal factor of the space-time geometry are encoded as properties of the quantum operator algebra.  {\it The space-time geometry is not a fluctuating quantum variable, but, in general, only a thermodynamic/hydrodynamic property of the quantum theory.}  This fits well with a prescient and beautiful paper by Jacobson\cite{tjentropy}\footnote{Jacobson's work has been followed by many papers by Padmanabhan and more recently by Verlinde\cite{padverl}. The latter work has led to an explosion of papers on entropic gravity.}, but not with decades of theoretical studies of quantum gravity as a quantum field theory.  The latter approach seems to have been validated by string theory.  However, I will argue that the appearance of quantized fields, and in particular of quantized fluctuations of the gravitational field, as exact concepts in string theory, is a consequence of the existence of scattering states localized on the boundaries of infinite, asymptotically flat or AdS geometries, combined with the usual connection between particles and fields.  By contrast, when we study finite causal diamonds, we find that they can contain only a finite number of particles, whose degree of localization is constrained.  More general states of the system correspond to black holes.  I'll demonstrate this by purely kinematic arguments, once we establish what the real fluctuating variables of QG are.   See the section on SUSY and the holographic screens.

\subsection{The Hilbert spaces of time-like trajectories}

In quantum mechanics, an observer is a large quantum system with many semi-classical observables.   A semi-classical observable takes on the same value in an exponentially large number $e^V$ of states, where $V$ is the size of the system in microscopic units.  Its time evolution is approximately classical, with corrections of order $1/V$ . The time evolution operator mixes those states on a time scale much shorter than classical motions of the observable.  This guarantees a washout of phase coherence between different classical trajectories, which is exponential in $V$.  In nature, there are two kinds of systems with such observables: systems of particles with local interactions, and black holes.  Black holes have many fewer classical observables, and many more quantum states, for a given size.   Any observer follows a time-like trajectory through space-time.

A time-like trajectory gives rise to a nested sequence of causal diamonds, corresponding to larger and larger intervals along the trajectory.  The holographic principle and causality postulates say that the quantum mechanical counterpart of this sequence is a sequence of Hilbert spaces, each nested in the next as a tensor factor.  The dynamics of the system must respect this tensor factorization.   There are two general kinds of space-time that we will consider.   Big Bang space-times begin at a finite point in time.  Each causal diamond is thought of as having its past vertex on the Big Bang hypersurface, and its future vertex one Planck time to the future of the previous diamond.  The dynamics is time dependent Hamiltonian dynamics, with a Hamiltonian that, at any time, couples together only degrees of freedom in the causal diamond up to that time.  As time goes to infinity, the size of the Hilbert space approaches a maximum dimension space ${\cal H}_{max}$, whose dimension might be infinite.   The time dependent Hamiltonian at any given time is a sum $H_{in} (t) + H_{out} (t)$, where $H_{in}$ operates in ${\cal H}_t$ and $H_{out}$ in the tensor complement of ${\cal H}_t $ in ${\cal H}_{max}$.  Note that there is nothing singular about the Big Bang in this formalism, it is just the point where the particle horizon shrinks to some minimal size. The familiar inverse power scaling laws of conventional cosmology arise as late time scaling behaviors in the limit of large particle horizon.

For time symmetric space-times we extend each causal diamond symmetrically around some space-surface of time symmetry.  The dynamical law is encoded in a sequence of transition operators between the past and future boundaries of each diamond.  In the large diamond limit, in asymptotically flat space, these converge to the scattering matrix.  In asymptotically AdS space, the areas of causal diamonds become infinite at finite proper time separation.  After that point, the boundary of each diamond contains a finite time-like segment of the infinite boundary.   The usual Hamiltonian evolution of AdS/CFT, obviously refers to evolution along this time-like segment, and its relation to the approximate scattering matrices of local observers is incompletely understood\cite{joepgiddings}.

The time evolution operator for a single observer is a complete description of all observations that observer makes over the entire history of the universe.  The key feature that makes HST a theory of SPACE-time, is its insistence on redundant descriptions of the same information by other observers.   That is, it consists of a whole set of nested sequences of Hilbert spaces and Hamiltonians\footnote{In order to avoid the awkward necessity of saying both "Hamiltonian" and ``partial S-matrix" at every turn, I'll henceforth assume we're talking about a Big Bang space-time.}.  Label the different sequences by an index ${\bf x}$.   Then, at each time $t$ and for each pair $({\bf x,y})$ we specify an {\it overlap Hilbert space} , $O(t;{\bf x,y})$ , which is a tensor factor of both ${\cal H} (t, {\bf x})$ and ${\cal H} (t, {\bf y})$.   The fundamental dynamical consistency condition of HST is that for every overlap, the density matrix implied by the time evolution in ${\cal H} (t, {\bf x})$ is unitarily equivalent to that implied by the time evolution in ${\cal H} (t, {\bf y})$.  This is both an infinite number of relations between the different Hamiltonians and a restriction on the initial state in each ${\cal H}_{max} ({\bf x})$.  

Below, we will introduce the {\it pixel Hilbert space} ${\cal P}$, such that each of the ${\cal H} (t, {\bf x})$ is a tensor power of ${\cal P}$.  We will insist that for each trajectory, there be a collection of {\it nearest neighbor trajectories} such that, at every time, the overlap between any pair of nearest neighbor trajectories is missing just one factor of ${\cal P}$.  We insist that the number of nearest neighbors be the same for every ${\bf x}$.  This makes the space of trajectory labels into a topological space of fixed dimension.   We are thus defining the topology of a space-like surface, through which all of our trajectories pass.   The dimension of the overlap Hilbert spaces for non-nearest neighbor points is required to be a non-decreasing function of the minimal number of steps between the two points.   Specifying the overlaps is equivalent to defining a discrete approximation to a Lorentzian geometry.  The individual trajectories are lines of fixed spatial coordinate in some coordinate system.   The areas and intersections of the causal diamonds can be read off the Hilbert space construction and re-interpreted, for large Hilbert space dimension, using the Bekenstein-Hawking formula.  In all of the models that have been studied so far, the space of trajectories is taken to be a hypercubic lattice.   However, the overlap rules depend only on the minimal number of steps on the lattice, and thus become rotation invariant for large numbers of steps.

To summarize: HST is an infinitely redundant description of the quantum system seen by a particular time-like observer.  The redundancy is described in terms of an infinite number of other observers, and the consistency conditions between them is a powerful constraint on the dynamics of the system.  Very few solutions of these constraints are known.  Space-time emerges from a purely quantum construction.  In the examples studied so far, Jacobson's principle that Einstein's equations emerge from the thermo-hydrodynamics of a system that obeys the Bekenstein-Hawking area law, is verified by explicit computation (see below).   The metric of space-time is encoded in the relations between various quantum Hilbert spaces and is not itself a fluctuating quantum variable.   The question of what the quantum variables are is answered in the next section.

\section{SUSY and the Holographic Screens}

Consider a {\it pixel} (a concept we'll define better in a moment), on the holographic screen.  It consists of a null direction pointing through the screen, and a bit of transverse hyperplane to that direction.  As emphasized by Cartan and Penrose, that directional information is encoded in a spinor satisfying
$$\bar{\psi} \gamma^{\mu} \psi (\gamma_{\mu} )_{\alpha}^{\beta} \psi_{\beta} = 0.$$  The spinor bilinear must be a null vector, and the equation says that $\psi$ is a null plane spinor for that vector, which means that all the other spinor bilinears lie in some transverse hyperplane.   The equation has local Lorentz symmetry and local rescaling invariance on the screen.  The first of these gauge symmetries can be fixed, by insisting that the null direction is determined by the coordinate on the sphere.  For example, if the screen is a sphere, then the null direction at the point parameterized by the unit vector ${\bf \Omega}$ is $(1, {\bf \Omega})$.  Thus, the variables are sections of the spinor bundle over the screen.   The quantum commutation relations will break the local scale invariance to a local $Z_2$, which is identified with $(-1)^F$.   The choice of a section of the spinor bundle determines a $d-2$ plane at each point on the screen.  Local Lorentzian geometry is built, in the vicinity of the screen by supplementing this hyperplane with the ingoing and outgoing null directions to the screen.

The naive notion of pixel can be thought of as a finite dimensional approximation to the algebra of functions on the screen.  The finite dimensional algebra has a basis consisting of characteristic functions which are equal to one on a single pixel and vanish elsewhere.  More general, perhaps non-commutative, approximations are known and the non-commutative ones are called {\it fuzzy geometries}.   The holographic principle in fact demands that the degrees of freedom on the holoscreen are finite in number.  The most elegant way to impose this condition is to put a sharp cutoff on the spectrum of the Dirac operator on the holographic screen\cite{tbjk}.  Since a finite area screen is a compact Euclidean manifold, the Dirac operator has a discrete unbounded spectrum, whose degeneracy at large eigenvalue $p$, scales like $|p|^{d-2}$. 
It is well known\cite{connes}, even to physicists, that the eigenvalues and eigensections of the Dirac operator completely determine the geometry of a manifold.  For example, the short time behavior of the heat kernel of the square of the Dirac operator satisfies

$$\langle x | e^{- t D^2} | y \rangle \sim t^{-\frac{d - 2}{2}} e^{- \frac{d^2 (x,y)}{2t}} ,$$ and therefor encodes both the dimension of the holoscreen and the geodesic distance function $d(x,y)$.  

A simple compactification of the HST formalism to a space of the form $K \times M^d$, where the second factor is $d$ dimensional Minkowski space, takes the spinor bundle to be a tensor product of the spinor bundle on the $d-2$ sphere and that on $K$, with separate eigenvalue cutoffs on the two Dirac operators.  The scaling of the degeneracies tells us that the radius of the sphere scales like $N$, the cutoff on the sphere's Dirac operator.  To understand this better, lets consider the case $d=4$ .  The cutoff Dirac spectrum on the two sphere contains all half integer angular momenta up to some maximum.  The chiral spinor is
an $N\times (N+1)$ matrix and its conjugate is an $(N+1)\times N$ matrix.  The quantum anticommutation relations take the form

$$[\psi_{i\ P}^A , \psi^{\dagger\ j}_{B\ Q}]_+  = \delta_i^j \delta_B^A Z_{PQ} ,$$
with commutation relations between $Z$ and $\psi$, which make this into a superalgebra with a finite dimensional unitary representation.  $P,Q$ label a basis of sections of the cutoff spinor bundle over $K$.
$Z_{PQ}$ lives in the corresponding cutoff bundle of forms.  If the manifold has a covariantly constant spinor, there is a zero mode of the Dirac operator, which is preserved by the cutoff.  In that case there is a unit operator contribution to $Z_{PQ}$, namely $Z_{00}$.  

The entropy of this system is $N(N+1) {\rm ln}\ {\cal P}$, where ${\cal P}$ is our symbol both for the Hilbert space of the irreducible representation of the superalgebra for fixed matrix indices, and its dimension.  The dimension scales like the exponential of the number of fermionic generators.   When the cutoff $P_K$ on the internal Dirac operator is large, ${\cal P} \sim P_K^{d_K}$, which scales like the volume of $K$ in higher dimensional Planck units.  These formulae define the four dimensional Planck mass $M_P$ and the higher
dimensional Planck mass $M_{4 + d_K}$, via
$$\pi (RM_P)^2 \equiv N^2 {\rm ln}\ {\cal P} \equiv \pi R^2 V_K M_{4 + d_K}^{2 + d_K} .$$

If we look for a compactification to $4$ dimensional Minkowski space with internal dimensions fixed in Planck units, we want to take $N\rightarrow\infty$ with fixed $P_K$, and we have to insist on conformal invariance on the resulting continuous $S^2$, since this is the Lorentz group $SO(1,3)$.  The conformal Killing spinor equation on $S^2$
   $$D_m q = e_m^a \gamma_a p ,$$ is conformally covariant and its solutions transform as right and left handed Weyl spinors under $SO(1,3)$.  When we take the $N\rightarrow\infty$ limit, we can find linear combinations of $\psi_i^A$, which converge to delta functions concentrated at points on the sphere $\psi \delta (\Omega - \Omega_0) ,$ where $\Omega$ and $\Omega_0$ are unit 3 vectors, and, 
   $$[\psi , \psi^{\dagger}]_+  = p.$$  $p$ is a positive normalization constant, which appears in taking the limit. If we smear these with the conformal Killing spinors, we obtain operators, which obey the commutation relations
   $$[Q_{\alpha}, Q_{\dot{\beta}}^{\dagger} ]_+ = \sigma^{\mu} P_{\mu},$$
   $$P_{\mu} = p (1, \Omega_0).$$
   These are the commutation relations for a single massless superparticle.
   
   Before discussing multi-particle states we note two features of this construction.  Poincare invariance arises as a consequence of super-Poincare invariance.  HST cannot describe asymptotically flat space without SUSY.  Secondly, the compact dimensions have {\it no moduli}.   The number of Dirac eigensections below some cutoff changes discontinuously with the moduli.  The quantum theory depends only on the number of eigensections, and their commutation relations form a superalgebra with a finite dimensional unitary representation, and have no moduli either. We can obtain approximate moduli only when $P_K \gg 1$, when various ratios of integers become almost continuous variables.  The ubiquity of moduli in stringy constructions of compact manifolds is an artifact of approximations in which some length scale is much larger than the Planck scale.  The idea of fixing the moduli with a potential is fundamentally flawed.  They are intrinsically discrete variables
   
   \section{Multiparticle States and Horizon States From Matrices}
   
   {\bf This is probably the most important section of this summary.}  To obtain multiparticle states, we partition the matrix $\psi_i^A$ into a block diagonal\footnote{What we really mean by this is that the square matrices $\psi \psi^{\dagger}$ and $\psi^{\dagger} \psi$, are block diagonal.}, with blocks of size $N_i$, $\sum N_i = N$, assume that the Hamiltonian decouples these blocks from the other variables, and take the limit of the previous section with the ratios $N_i / N_j $ fixed.   The permutations of the blocks are just relabeling or gauge transformations, which we interpret as particle statistics.  The correlation between anti-commutation and half integral spin, ensures the usual correspondence between spin and statistics.  Considering all possible such decompositions as $N\rightarrow\infty$, we obtain a Fock space of super-particles.  The Hamiltonian, as always in scattering theory, is just the sum of single particle Hamiltonians.
   
 To describe scattering we must recognize that there was a parity ambiguity in our identification of conformal Killing spinors, and we equally well have constructed SUSY generators that anti-commuted to $$p (1, - \Omega),$$ the outgoing null vector\footnote{This is also the key to obtaining descriptions of massive particles.  The incoming momentum of a massive particle is written as the sum of an incoming and outgoing null momentum.}.   Thus, asymptotically flat space is described by two Fock spaces, one of ingoing and one of outgoing particles.   The scattering matrix is the map between them.  The biggest lacuna in the theory of HST is the absence of a set of equations which determine the S-matrix.  I will discuss this briefly in the conclusions.
 
 A natural question at this point is what the role of the rest of the matrix variables is.  The answer to this depends on whether we are interpreting our system as simply a finite area causal diamond in Minkowski space, or the entire causal patch of dS space.  Let us start with the first interpretation.   Using the formalism of QFT, Unruh\cite{unruh} showed that an accelerated observer sees the no-particle state in Minkowski space as thermal state, with an acceleration dependent temperature.  We want to find a description of the Unruh effect in the HST formalism.  The Hamiltonian of HST is observer dependent and it's obvious that we should think of the SUSY algebra we have just constructed, with its associated Hamiltonian, as the Hamiltonian of the unique geodesic observer in our causal diamond.  I will only sketch the ideas for constructing the Hamiltonians of accelerated observers, which will be more fully developed in \cite{tbacc}.
 
 The SUSic Hamiltonian is bilinear in fermion operators, and decouples the block diagonal variables from the rest.   Consider adding other single trace operators to it.   In a recent paper, Susskind and Sekhino pointed out\cite{sussfastscramble}, that horizons are fast scramblers of information, perhaps the fastest in nature.  They scramble information in a time logarithmic in the number of degrees of freedom.  They further conjectured that generic quantum Hamiltonians built from traces of $N\times N$ matrices are fast scramblers.
 Imagine then that we modify the SUSIC Hamiltonian to the form
 $$H = Z H_{SUSY} + \frac{1}{N^2} {\rm tr}\ f (\psi^{\dagger}\psi ) .$$  We have suppressed the indices $P,Q$ describing the internal spinor bundle.  $f$ is a general polynomial with order one coefficients in all possible fermion bilinears\footnote{The discussion is easily generalized to include multi-trace operators.  We can write these in terms of single trace operators coupled to auxiliary variables, which have no kinetic energy.  In the large N limit, fluctuations of the auxiliary variables are suppressed and we obtain the same physics as a self consistent single trace Hamiltonian, up to small corrections.}. Note that we have chosen the coefficient of the non-linear terms such that the 't Hooft couplings are $o(1/N)$.  For the geodesic observer, $Z=1$ and the system will thermalize at a temperature of order $1/N$.   For accelerated observers, we simply choose $Z $ to be a decreasing function of $a$, such that the temperature increases linearly with the acceleration.  When the acceleration is $\sim N$, the temperature is Planck scale and all of the $N^2$ degrees of freedom are rapidly scrambled. There is no longer any distinction between particles and the rest of the system.
 
 We can also do a simple calculation of the maximal field theoretic entropy one can have in a finite causal diamond.  High entropy states in QFT come from many particles with maximum momentum.  In HST, the momentum plays the role of an angular momentum cutoff, and we need high momentum in order for a particle to be approximately localizable on the sphere.  This conflicts with the many particle criterion, because $\sum N_i = N$.  The obvious compromise is $N_i \sim \sqrt{N}$, which for large $N$ allows us to have a maximal number of highly localized particles.  This gives a total entropy of order $N^{3/2}$ , which is the same scaling one obtains by considering the field theoretic states in a finite, roughly spherical, volume, whose gravitational back-reaction does not produce a black hole whose size scales like the radius of the region.  This dynamical gravitational effect is entirely incorporated in the kinematic entropy counting of the states of the HST formalism.  This is the sense, in my opinion, in which ``gravity is an entropic force"\cite{verlinde}.  It is of course natural to identify the rest of the degrees of freedom of the system with the states of ``a black hole in the causal diamond".  Readers familiar with the teleological definition of a black hole in general relativity will not take kindly to this suggestion, but they should be reminded that we are doing quantum, rather than classical, gravitational physics. What I mean precisely by my statement is that, if the Hamiltonian of the system in this causal diamond is such that the states are being constantly scrambled at the fast scrambler rate, then, when viewed from the perspective of a much larger causal diamond, this state will behave like a black hole.   Indeed, this is almost inevitable, if the dynamics of the large causal diamond is described by a Lorentz invariant S-matrix.  More details of this assertion will be supplied in \cite{tbacc}.   
 
 The momentum of individual particles in this scheme, is proportional to block size, as in Matrix Theory\cite{bfss}. If we take the unit of momentum to be $1/R$, then the UV cutoff is of order $(\frac{M_P}{R})^{1/2}$, which agrees with the field theoretic estimate.  In QFT this very low cutoff is somewhat puzzling.  We know we can make states of much larger momentum, but there does not seem to be a nice way to implement the no black hole constraint directly in QFT.  By contrast, in the HST formalism, it is clear that we can have particles of higher momentum at the expense of having fewer of them, with less entropy.  If we insist on the maximal angular localization consistent with the overall momentum, then the highest momentum we can get comes from having $o(1)$ particles described by $N^{3/4} \times N^{3/4} $ matrices.  For $R$ of order our current horizon radius, this is the TeV scale.   We can get fuzzier particles, with higher momentum, by considering states with each $N^{1/2} \times N^{1/2}$ block in {\it the same} angularly localized state.   This gives momenta of order the Planck scale as the maximum.  The smaller degree of angular localization for particles above the TeV scale is interesting, and might have experimental consequences. Note however that for $R$ of order our current horizon volume, $N^{1/2} \sim 10^{30}$ so the apparent lack of angular precision predicted by the theory is probably not measurable.  All of this talk of our current horizon radius suggests that we should be moving on to the theory of dS space.
 
 \subsection{Particles and Black Holes in dS Space}
 
 The discussion above can be taken over directly into dS space.   In this case we think of $N$ as fixed and finite.  An interesting feature of the formalism, which has no obvious interpretation in the finite causal diamond in Minkowski space model, is that the bilinear Poincare Hamiltonian actually describes $\sim N^{1/2} $ copies of the particle states.  In our model of dS space, we interpret these as particles in $N^{1/2}$ causally disconnected horizon volumes.  A generic state of the system will not be a tensor product of particle states in the different horizon volumes.  According to the HST theory of dS space\cite{tbfm} the maximally uncertain density matrix of the full system is identified with the dS vacuum and the Gibbons-Hawking entropy is the logarithm of the dimension of the full Hilbert space.   Consider a state with a number of particles of order $1$ and a total momentum $\ll M_P$.   These are represented by a matrix with a small number of blocks whose size sums up to $\sum N_i \ll N$.   The entropy of the horizon degrees of freedom is lowered by decoupling these blocks from the system, by an amount proportional to
$$N^2 - (N - \sum N_i )^2 - \sum (N - N_i) N_i \sim N \sum N_i .$$   In other words, the entropy of a particle state in a single horizon volume is smaller than that of the maximally uncertain density matrix by an amount linearly proportional to the energy of the particles.  This means that, from the point of view of the Poincare Hamiltonian, the maximally uncertain density matrix is thermal.  The coefficient of $N$ in front of the entropy deficit implies that the temperature is inversely proportional to the dS radius.   
 
 We describe black holes in dS space by a similar strategy.   Our $o(1)$ particle states occupy part of a block of size roughly $\sum N_i \times \sum N_i$ of the matrix variables.  Black hole states are defined by the constraint
 $$\psi_{i\ P}^A | BH \rangle = 0,$$ where $i = 1 \ldots N_- ,$ and $A = 1 \ldots N_+$.   $N_-$ is an integer less than $N/2$ and $N_+$ is the closest integer to the solution of the quadratic equation $$N^2 = N_+^2 + N_-^2 + N_+ N_- .$$ 
 Note that the ratio of the entropy deficit of these configurations to the total dS entropy is $N_+N_- / N^2$ , which is the formula for Schwarzschild-dS black holes.   For such black holes the formula for the mass in Planck units is
 $$M = \frac{1}{2R^2}(R_+ R_- (R_+ + R_-)) =  \frac{1}{2}(1 - \frac{N_+^2 + N_-^2}{N^2} (N_+ + N_-)\sqrt{\frac{{\rm ln}\ {\cal P}}{\pi}} .$$ In \cite{tbfm} we exhibited an operator $P_0$ whose statistical average in the black hole states reproduced this formula for large $N$, with fluctuations of order $1/N$.
 
 To summarize: the holographic theory of dS space, like that of a finite causal diamond in Minkowski space, has a one parameter set of Hamiltonians, representing physics as seen by observers following, {\it e.g.} trajectories of fixed static coordinates (Rindler coordinates for Minkowski space).  Each Hamiltonian consists of a bilinear piece and a small multilinear piece in the fermionic pixel variables. For large dS radius, $N$, in Planck units, the bilinear piece approaches $N^{1/2}$ copies of the Hamiltonians of $\leq o(N^{1/2})$ massless superparticles.  The coefficient of the bilinear piece is a redshift factor, which controls the temperature at which the multilinear piece thermalizes these particle states.  It is $T =  (2\pi R)^{-1}$ for the geodesic observer, and approaches infinity for the maximally accelerated observer. This formalism explains the thermal nature of the dS vacuum and its unique temperature, as well as the gross properties of both particles and black holes.
 
 \section{Holographic Cosmology}
 
 This section summarizes a long series of papers written with W. Fischler and other collaborators\cite{holocosm}.  The basic idea is that a Big Bang cosmology, from the point of view of any given observer is a time dependent Hamiltonian
 (actually a discrete time evolution operator), which couples together only  $N(N+1) {\rm ln}\ {\cal P}$ degrees of freedom at cosmological time $N$.  The operators satisfy the basic super-algebra of HST,
 $$[\psi_{i\ P}^A , \psi^{\dagger\ j}_{B\ Q} ]_+ = \delta_i^j \delta_B^A Z_{PQ},$$ with a {\it fixed fuzzy compactification}.
 The Hamiltonian $H(N)$ is chosen, independently at each time, from a random Gaussian distribution, with the proviso that as $N\rightarrow\infty$ it approaches the Hamiltonian of a $1+1$ dimensional CFT, with central charge of order $N^2$, with cutoff of order $N$ on an interval of length $1/N$.   It also contains a random irrelevant perturbation of this field theory.   The average energy is of order $N$, which is what one would expect from a horizon filling black hole, and the energy density $\rho = \langle H(N) \rangle / N^3$ is $\sim 1/N^2$, which is what one expects from an FRW universe.  The entropy density is $\sigma \sim 1/N \sim \sqrt{\rho}$.  This is the relation for a perfect fluid with equation of state $p=\rho$.   This is consistent with the fact that the model saturates the covariant entropy bound, and Fischler and Susskind showed that the only FRW model that can do that is the flat $p=\rho$ universe\cite{fs}.
 
 The catalog of observers for this model is an infinite cubic lattice.  The causal diamond Hilbert space for each observer at time $N$ is ${\cal H} (N, {\bf x}) = {\cal P}^{N(N+1)} $, and the overlaps are ${\cal O} (N, {\bf x,y}) = {\cal P}^{N({\bf x,y}) (N ({\bf x,y}) + 1)} $, where $N({\bf x,y}) = (N - d({\bf x,y})) \Theta [N - d({\bf x,y})] $.  $d({\bf x,y})$ is the minimal number of lattice steps between the two points. The time evolution operators and initial state are identical at all ${\bf x}$, and this is required in order to satisfy all of the consistency conditions.
 The model thus has an emergent space-time, which is homogeneous and isotropic and flat for generic choice of initial conditions. Isotropy appears in two different ways.  The Hamiltonian of each observer is required to be invariant under rotations acting on the pixel variables, which are sections of the cutoff spinor bundle on the two sphere.  The causal distance defined by the overlap rules becomes rotation invariant in the large $N$ limit.  The locus of all points a fixed number of steps from a given point is a cube tilted at $45$ degrees to the lattice axes.  According to the overlap rules, these points are all at the same distance in the physical metric, so the geometry of this cube is spherical in the large $N$ limit.  
 
 This model is called the dense black hole fluid (DBHF) because it has the thermodynamic characteristics of a collection of black holes, which are at each instant, within a Schwarzschild radius of each other, and which constantly merge so that the particle horizon is filled with a black hole saturating the covariant entropy bound.  One should not take this intuitive picture too seriously.  The model is well defined mathematically and has the coarse grained properties of the flat $p=\rho$ FRW.   If the reader finds the DBHF terminology confusing he/she is free to ignore it and concentrate on the mathematics.
 
 \subsection{Holographic Eternal Inflation, and the dS Black Hole in the $p=\rho$ Background}
 
 We now introduce two variations on the DBHF model, which are ingredients in the construction of a realistic cosmology.   In the first, called Holographic Eternal Inflation (HEI), we simply stop the growth of the Hilbert space in the DBHF model at some fixed cosmological time $n$, and rescale the cutoff $M$ and the interval $L$ on which the $1+1$ CFT is defined, so that $ M \sim 1/n$ and $L\sim n$, so that the new Hamiltonian has $\langle H_{new} (n)\rangle \sim 1/n $.   This rescaling is done gradually as the cosmological time approaches $n$.   We then continue the Hamiltonian evolution forever with this fixed Hamiltonian.   The late time evolution is thus identical to that of dS space, viewed in static coordinates by the maximally accelerated observer near the horizon.   The entire evolution is that of a flat FRW model, which begins with $p=\rho$ and evolves to $p= -\rho$ with a Hubble scale $n$.  The coordinate system is one in which the FRW coordinates morph asymptotically into the static time in dS space\cite{sussetal}.  
 
 The overlap rules of this new model are the same as the old one, which means that lattice points, which are more than $n$ steps apart, never have any overlaps.   Note that although the late time Hamiltonian evolution of any individual observer is identical to that of our model of dS space, this model has an infinite number of independent copies of the degrees of freedom of that model.  Indeed, if one chooses a point on the lattice, no point outside of the tilted cube with $n$ steps has any overlaps with it.  Thus, we can tile the lattice with an infinite number of disjoint tilted cubes, and declare that the degrees of freedom at the centers of each of these cubes are independent of each other.
 
 Our second new model uses just one tilted cube of the HEI model, and introduces new overlap rules:  No point on the interior of the tilted cube has overlap with {\it any} point outside the cube.   At points outside the tilted cube we use the full DBHF model, where the growth of each observer's Hilbert space is unbounded.  We modify the overlap rules so that there are no overlaps between interior points of the tilted cube and exterior points.  The overlaps with points on the boundary of the tilted cube can of course grow only to the maximal size of the Hilbert space on those points, namely
 ${\cal P}^{n(n+1)}$.   These new rules are completely consistent.
 
 The result is obviously a marginally trapped spherical surface embedded in the $p=\rho$ fluid.  Indeed, the Israel junction conditions allow for the existence of such a black hole in the $p=\rho$ FRW space-time, with a dS interior.
 At very late (exterior) FRW times, in the vicinity of the black hole horizon, these solutions look like the thin wall limit of the static solutions of Mazur and Mottola\cite{mazmott}.  This solution is classically stable, as a consequence of the familiar black hole and dS no-hair theorems.  If we constructed it in an effective theory of a scalar field coupled to gravity, we would conclude that it was unstable to Hawking radiation.  However, this claim makes no thermodynamic sense.  A conventional black hole embedded in flat space, or {\it e.g.} a radiation dominated FRW space-time, decays because it can increase its entropy by exciting the {\it assumed} adiabatic vacuum state of the exterior space-time.  However, the explicit $p=\rho$ model that we have constructed, is not in any sort of adiabatic vacuum state.  Instead, it's state varies over the entire causally accessible Hilbert space.  Its time averaged density matrix is maximally uncertain.  The underlying quantum model for the black hole with dS interior has no quantum instability.  {\it Thus, the effective quantum field theory analysis of this model is completely wrong, even though the model's thermodynamics is described by an effective classical field theory.}  This is, I would claim, a generic property of systems in QG, which are close to saturating the CEB.  We will discuss this further in the next section.
 
 In an eventual model of our own universe, the final state of this dS black hole model will be identical to our own final state.   The reader will have noted that the value of the dS Hubble radius $n$ is a free parameter in both of these models.  We can exploit this to obtain a framework for environmental selection of the c.c..   Imagine trying to construct a model which has some distribution of such dS black holes, of various sizes, sprinkled throughout the $p=\rho$ universe.  We do not have a quantum construction of such a model, but we can use what we know about Einstein's equations to understand how it would evolve.   Depending on initial conditions, some of the black holes will collide and merge to make larger black holes, while others will be driven apart by the Hubble expansion in the $p=\rho$ background.   We always imagine that the distribution is sparse enough that the black holes do not come to dominate the energy density and change the overall expansion rate.  For example, we could just have a large but finite number of black holes.   Notice that in HST, each different configuration of black holes will be a different quantum model.   If we now suppose that there are any sort of biothropic selection criteria that are affected by the value of the 
 c.c.\footnote{Because of the connection between the value of the c.c. and SUSY breaking in HST, the biothropic effects of the c.c. are drastically different than they are in other kinds of models.}, then we will be forced to a model that contains at least one dS BH with a c.c. in the biothropic range.
 
 \subsection{Effective Field Theory}
 
 {\it This is a very important section, because it proposes a drastic revision in the way effective field theory emerges from quantum gravity.}  In 1995, Ted Jacobson wrote a prescient paper\cite{tjentropy}, which demonstrated that Einstein's equations followed from the laws of thermodynamics for a Lorentzian space-time, associated with a system obeying the Bekenstein-Hawking connecton between area and entropy, for every maximally accelerated observer at every local Rindler horizon.  The stress tensor on the right hand side of the equations is not specified by Jacobson's argument, except that its integral must give the energy used in the laws of thermodynamics. HST defines space-time in terms of the Bekenstein-Hawking law, and so it should, and so far does, give rise to systems satisfying Einstein's equations.  Jacobson argued that the metric in Einstein's equations might be just a coarse grained thermo/hydrodynamic variable, in which case it might be inappropriate to consider it a quantized field.
 
 On the other hand, string theory in asymptotically flat and AdS space-times seems to have a much closer connection to quantized bulk field theory.   The S-matrix/boundary correlation functions of the exact theory of quantum gravity are, in certain approximations, computed from Feynman diagrams of a quantized field theory in the bulk.  HST gives us a way of understanding both the Jacobsonian point of view and that of conventional string theory.  Indeed, we have seen that the variables of HST in a large causal diamond can be separated into particle states well described by quantum field theory, and horizon states which cannot be associated with localized excitations in the bulk.   The asymptotically flat or AdS boundary conditions on bulk field theory, imply in particular that for asymptotically large causal diamonds, we only allow particle states.   Correspondingly, when we attempt to describe systems with such boundary conditions in HST, we must restrict attention to the Poincare Hamiltonian, which decouples the particle states from non-local states on the horizon\footnote{The analogous statement for asymptotically AdS space-time has not yet been worked out.}.
 
 It is well known that the low energy expansion of any unitary, relativistic, crossing symmetric S matrix can be described by an effective quantum field theory, so a theory of particles in asymptotically flat space-time must look like a quantum field theory at low energy.   By contrast, when we study situations in which the covariant entropy bound is nearly saturated, we know that quantum field theory breaks down, but Jacobson's argument shows us that we should still expect to see classical field equations describing the thermo/hydrodynamics of space-time.  These effective fields should NOT be quantized.  To distinguish the two situations, we should perhaps talk about Thermodynamic Effective Field Theory (THEFT) and Effective Quantized Field Theory (EQFT).  
 
 In general we should expect that an underlying HST model should give us equations that determine the stress tensor sitting on the RHS of Einstein's equations.  For the DBHF model, it is sufficient to specify the equation of state, but this is not the case for the other two models.  For those, we follow the moral equivalent of the extraction of an effective Lagrangian from the string S-matrix:  we find a simple field theory which reproduces the coarse grained physics of the underlying model.   For the HEI model this field theory is simply
 $${\cal L} = \sqrt{-g} [R - (\nabla\phi)^2 + V_0].  $$  The general flat FRW solution of this model has a scale factor
 $$a(t) = c \sinh^{1/3} (3 t H) ,$$ where $H$ is the Hubble constant corresponding to $V_0$ and is fit to the inverse Hubble radius of the underlying model.  The constant $c$ is fixed by requiring that in the $p=\rho$ phase the CEB is exactly saturated.
 
 The important point now is that the quantized fluctuations of this field theory have nothing to do with the underlying fluctuations of the HST model.  The latter are described in terms of generic scrambled states of all the $\psi_i^A$ variables, and there is no decoupling of particle degrees of freedom.   If the c.c. is small, one can introduce the Poincare Hamiltonian in the asymptotic future dS space, but the typical state predicted by this cosmology will be the empty dS vacuum with no localized excitations.  The Poincare Hamiltonian will have a zero eigenvalue in most of these states, with non-zero eigenvalues (Boltzmann particles) appearing only as occasional thermal fluctuations.   The effective metric and scalar fields are just a classical summary of the hydrodynamics of the underlying model.  In particular, $\phi$ has {\it nothing} to do with any of the quantized fields, which give an approximate description of particle scattering in the asymptotic dS future.  
 
 Similar remarks are appropriate for the holographic quantum cosmology whose coarse grained description is a dS black hole embedded in the $p=\rho$ model.  The effective field theory for this model again has an additional scalar field in addition to the metric.  The potential has a local minimum at $\phi = 0$, with a positive c.c. that is fit to the Hubble radius of the dS space.  At infinity the potential falls rapidly to zero and there are no other minima, nor negative potential regions. One restricts attention to spherically symmetric field configurations and chooses initial conditions such that
 \begin{itemize}
 \item There is a Big Bang singularity, near which the energy density is dominated by scalar kinetic energy and the equation of state is $p=\rho$. The constants are chosen so that the CEB is exactly saturated.
 
 \item Inside some coordinate radius $r_0$, the field is in the basin of attraction of the origin and it does not leave this basin of attraction during the kinetic dominated era.
 
\item There is a smooth radial transition to field values at $r > r_0$ which are in the basin of attraction of $r = \infty$.
At infinity, the field blows up logarithmically and its gradient goes to zero.  

\item It is clear that a black hole will develop in this model.  The positive c.c. in the interior invalidates the Hawking-Penrose singularity theorem.  We fine tune both the parameters of the potential, and the initial conditions, so that we get a non-singular solution where the black hole interior asymptotes to a static patch of dS space.  The fact that there are solutions of the Israel junction conditions with these properties, strongly suggests that such non-singular solutions
will exist for some potentials and initial conditions.   Our quantum HST model is completely consistent and clearly represents a spherically symmetric trapped surface with dS interior and $p=\rho$ exterior.

\end{itemize}

Spherically symmetric gravity with a single scalar field has no field theoretic degrees of freedom.   An effective field theorist would be tempted to insist that one could not neglect fluctuations violating spherical symmetry.  He/She would claim that this black hole was quantum mechanically unstable to Hawking radiation.  A Jacobsonian THEFTist would counter that in this situation, where the underlying cosmology saturates the CEB, effective fields play only a constrained role, and are classical variables describing the coarse grained thermodynamics of the underlying model.
Their Lagrangian must be tuned to reproduce the results of HST, and has no fundamental significance. Hawking radiation cannot occur because the space-time external to the black hole already saturates the CEB and has no particle excitations.  Black holes in flat space are unstable because entropy can increase when we emit particles into the pure Minkowski vacuum state.   The dS black hole in the $p =\rho$ background is thermodynamically stable.

\section{Towards a realistic cosmology}

This section records work in progress, whose aim is to combine the models above into a realistic cosmology.  The cosmology that evolves from a DBHF to a dS black hole\footnote{It's probably best to recall here that this phrase refers to a static patch of dS space joined to the horizon of a black hole in the $p=\rho$ FRW, not to a Schwarzschild-dS black hole.} with radius $N \sim 10^{61}$ embedded in the $p=\rho$ FRW is a broad brush picture of our universe if we make the natural assumption that its initial state is the DBHF and its final state a stable dS space.   However, if we concentrate on the Poincare Hamiltonian, which describes particles propagating in the asymptotic dS space, we find that it is extremely unlikely to find any.  The dynamics produces a typical state in the dS ensemble, and the overwhelming majority of those states look like the empty dS vacuum.  Localized ``Boltzmann excitations", will appear as low probability thermal fluctuations, and organized structures like living creatures will be even more rare.

Fischler and I\cite{holoinflation} propose to remedy this situation by introducing an intermediate stage of inflation, with a Hubble radius $1 \ll n \ll N$.  We will do this by taking $e^{3 N_e}$ of the independent horizon volumes in the HEI model, where $e^{3N_e} = (\frac{N}{n})^2 $, and then allowing the Hamiltonian of the HEI model to become time dependent again, in such a way that these independent degrees of freedom interact with each other ``in a way consistent with local bulk physics".   The Hamiltonian gradually converges to the Poincare Hamiltonian of the asymptotic dS space with radius $N$.   
 The quotation marks in the penultimate sentence indicate that we do not have a microscopic description of this transition, using the rules of HST.  However, it goes without saying that a time dependent Hamiltonian, which interpolates between the sum of $e^{3N_e}$ copies of the static Hamiltonian for dS space with Hubble radius $n$, and the Poincare Hamiltonian of the dS space with Hubble radius $N$ exists.  The real issue in question is one of understanding in depth how the holographic formalism can mimic locality.  Our argument for locality has so far been indirect, and utilized the folklore relating particle S-matrices to local Lagrangians at low energy.  Here we need a much more visceral demonstration of the way in which locality emerges.
 
 In \cite{holoinflation} we will argue that if such an interpolating Hamiltonian can be found, the resulting model will have the following properties:
 
 \begin{itemize}
 
 \item There will be approximately Gaussian fluctuations, of magnitude roughly $\frac{1}{n}$, which take on an approximately dS invariant form.  dS invariance follows from a discrete group, which converges to the dS group in the limit $N/n\rightarrow\infty$.  With our current limited understanding, we are unable to make any comments about the tilt of the spectrum, or the spectrum of gravitational waves.
 
 \item Inflation is very different from dS space.  There is only one horizon volume of the latter, many of the former.
 
 \item The origin of the fluctuations: The time averaged density matrix in each inflationary horizon volume is proportional to the unit matrix and the tensor product of these is just the dS vacuum density matrix. The fluctuations come from small variations, localized on a scale smaller than the inflationary Hubble scale. One way to think about them, which is reminiscent of the conventional picture, is that when the horizon volumes are allowed to interact, there are fluctuations in the local time coordinate.  Thus the pure state in each horizon volume is slightly different than in its neighbors, even though the time averaged state is the same.
 
 \item The fluctuations involve ALL of the states of individual inflationary horizon volumes, not just those describable by local field theory in a horizon.  The expansion of the universe after inflation gives "room" for many of these states to be realized as field theory fluctuations in the late time dS space.
 
 \item Reheating MUST occur, because the excitations of the asymptotic Hamiltonian are just multi-particle states.  However, we do not have a description of the process in microscopic terms.
 
 \item Jacobson's observations about the laws of thermodynamics and field equations, show us that there must be an effective field theory of this model.   The only obvious candidate is a slow roll inflation model with a single field.   As in our other models, this inflaton field is not a quantized field and is not to be identified with the fields of asymptotic particles in the late time, almost Minkowski, universe.
 
 \end{itemize}
 
 The details of these arguments will be presented in \cite{holoinflation}.
 
 \section{Conclusions}
 
 Holographic space-time is a generalization both of field theory and of string theory and can describe both particle excitations and regimes where the CEB is nearly saturated.  In the latter situation, effective field theory has only a thermodynamic meaning and the true quantum dynamics of the system is not well described by quantum fluctuations of the effective fields.  We've argued that the inflaton field is likely to be such a purely thermodynamical variable.
 
 The connection to field theory is obvious.  Field theory encodes the causal structure of space-time in the commutation properties of a net of operator algebras. HST simply specifies the, generally finite, dimension of those algebras, and uses it to encode the conformal factor of space-time.  Hamiltonian dynamics is very different in HST and QFT.  In QFT we have a single Hamiltonian, given formally as an integral over operators localized at points.  In HST there is an infinite collection of (generally time dependent) Hamiltonians, identified with an infinite collection of non-intersecting time-like trajectories, densely packed in space-time.  It is an inherently observer-centric formalism and none of the observables are ``generally coordinate invariant", except in the sense that they are operators referred to a particular physical gauge.
 The fundamental principle of HST is that pairs of observers, which share information, must give rise to unitarily equivalent density matrices on common tensor factors in their Hilbert spaces.
 
 I have sketched the way in which particles emerge as a subset of the matrix degrees of freedom describing a causal diamond and the construction of a one parameter set of Hamiltonians, related by rescaling the free particle part of the Hamiltonian, which capture the Unruh effect.  This formalism also has applications to particles and black holes in dS space, and reproduces a variety of gravitational formulae.   In particular, it realizes the idea that the dS vacuum is the maximally uncertain density matrix in quantum dS space and that the unique temperature of dS space is explained in terms of the degeneracies of the Poincare Hamiltonian, which describes particles.  The general way in which particles and quantized effective fields arise from this formalism makes it clear that the use of effective field theory ideas like vacuum state and vacuum tunneling have very limited applicability in THIS theory of quantized gravity.  It's also clear why effective quantum field theory {\it is} useful in asymptotically flat and AdS space.  Those boundary conditions ensure that there are an infinite number of particle states, which are decoupled from most of the excitations in large enough causal diamonds.
 
 The connection between HST and string theory comes via supersymmetry.  The HST formalism constructs the momentum operator from the supercharges, so it never has Poincare invariance without super-Poincare invariance.  Much of string theory can be derived from SUSY, with strings appearing as wrapped BPS 2 or 5 branes in a variety of geometries.  The HST analog of a compactification is a factorization of the spinor bundle over the holographic screen into the product of a spinor bundle over a $d-2$ sphere and that over some compact manifold $K$, with independent eigenvalue cutoffs on the Dirac operators on these factor bundles.  The question of whether the compactification is AdS or Minkowski, depends on the way in which we take the limit as the cutoff of the sphere's Dirac operator is taken to infinity.  Supersymmetry is preserved if the internal manifold has one or more covariantly constant spinors, which correspond to zero modes of the Dirac operator\footnote{There is a simple generalization of the Dirac operator to the case of flux compactifications.}.  This is necessary in order to obtain a Poincare invariant limit.
 
 In general, HST compactifications contain no moduli.  Continuous parameters arise as ratios of large integers, when we take the cutoff on the internal Dirac operator to infinity.  This corresponds to taking a large volume for the internal manifold.  I claim that moduli and the moduli problems are artifacts of taking limits where things are calculable, when some length scale is much larger than the Planck scale\footnote{The case of lines of fixed points in AdS models needs a separate discussion, but is by definition a situation in which infinite limits have been taken.}.
 
 Let me end by outlining the potential consequences of these ideas for phenomenology.  The fact that there are no moduli eliminates a whole set of cosmological models and problems, and the possibility that quantum gravity provides a natural candidate for the QCD axion.   Inflationary models are changed profoundly, in ways sketched above.  However, the most dramatic effect if the prediction of a very low scale for supersymmetry breaking.  I've not described the argument above because it is on a somewhat less solid footing than the ideas I've discussed.  It leads to the estimate $$m_{3/2} = K \Lambda^{1/4} ,$$ with $K$ a constant of order $10$.  The value of $K$ reflects the relatively large volume of the internal manifold, which is required to explain\cite{witstrong} the ratio between the Planck scale and the unification scale.  Using conventional supergravity formulae, this leads to a maximal scale of SUSY breaking
 $$F \sim (30 TeV)^2 .$$   In order to obtain acceptably large gaugino masses, the chiral field whose $F$ term breaks SUSY must be coupled to the standard model gauge fields through dimension $5$ operators scaled by $M \simeq $ TeV.  This low scale can only be generated by a new set of strong interactions, with the standard model acting as part of the flavor group of that model.  As a consequence, we have modifications of the running of standard model couplings from very low energies.  To date, the only class of models compatible with these ideas, and with perturbative gauge coupling unification, are the {\it Pyramid Schemes}\cite{pyramid}.  These models use Glashow's trinification, which embeds the standard model in $SU(3)\times SU(3) \times SU(3)$. They introduce a new $SU(N)$ gauge group with $N = 3$ or $4$, and three sets of chiral fields $T_i $ in the $(N, \bar{3}_i ) + (\bar{N}, 3_i)$.   In order to have the possibility of SUSY breaking, we must also introduce three singlets, $S_i$, which break SUSY through the O Raifeartaigh mechanism.  The only plausible candidate for dark matter (and only for $N=3$) is a hidden sector {\it pyrmabaryon}, a standard model singlet cubic in one of the $T_i$ fields.  This can't be a thermal WIMP, but it can be the dark matter if an appropriate asymmetry is generated in the very early universe.  This opens up a variety of scenarios for relating the dark matter density to the ordinary baryon asymmetry.
 
 Interestingly, in order to avoid Landau poles in the new gauge coupling for $N=3$, we must add renormalizable superpotential terms that violate at least two of the hidden sector baryon numbers.  The resulting dark matter candidate has either electromagnetic or chromomagnetic dipole moments and this can change its cross sections on nuclear targets and, in the electromagnetic case, might have important astrophysical consequences.
 
 The bottom line is that, if the scaling law for SUSY breaking is correct, the theory of Holographic Space Time gives us a highly constrained model for TeV scale physics, which has visible and quirky consequences at the LHC and in dark matter detection experiments.


\end{document}